\documentclass[preprint,showpacs,preprintnumbers,amsmath,amssymb]{revtex4-2}
\usepackage{graphicx}
\usepackage{dcolumn}
\usepackage{dcolumn}
\usepackage{bm}
\newcommand{\mbx}[1]{\mbox{\boldmath $#1$}}
\begin{document}
\preprint{AIP/123-QED}
\title{Statistical Mechanical Analysis of Gaussian Processes}
\author{Jun Tsuzurugi}
\email{juntuzu@ous.ac.jp}
\altaffiliation{
	The Department of Information Science and Engineering, 
	Okayama University of Science.}
\date{\today}
\begin{abstract}
In this paper, we analyze Gaussian processes using statistical mechanics. Although the input is originally multidimensional, we simplify our model by considering the input as one-dimensional for statistical mechanical analysis. Furthermore, we employ periodic boundary conditions as an additional modeling approach. By using periodic boundary conditions, we can diagonalize the covariance matrix. The diagonalized covariance matrix is then applied to Gaussian processes. This allows for a statistical mechanical analysis of Gaussian processes using the derived diagonalized matrix. We indicate that the analytical solutions obtained in this method closely match the results from simulations. 
\end{abstract}
\maketitle
\section{Introduction}\label{Introduction}
With the publication of seminal texts on machine learning\cite{Bishop}, it can be said that the field of machine learning is nearly complete.
Within the field of machine learning, there exists the Gaussian process\cite{Williams-Gau}-\cite{MacKay(2003)}. While there are already studies that have analyzed Gaussian processes using statistical methods\cite{Cressie(1993)}, no research has yet solved Gaussian processes by deriving exact solutions.
In this paper, we aim to perform a rigorous analysis of the Gaussian process using statistical mechanics.

To perform a statistical mechanical analysis, we enable mathematical calculations through modeling\cite{Gau-juntuzu1}. Specifically, we first impose periodic boundary conditions. Next, although the inputs of the Gaussian process are multidimensional, we consider them to be one-dimensional. These model assumptions facilitate discussions in Fourier space and allow for the diagonalization of the covariance matrix, i.e., the kernel matrix.

In this study, we focus on the case where the input points of the Gaussian process are equally spaced, and we derive an analytical solution for the mean squared error using a statistical mechanics approach. Here, we discuss the validity and significance of assuming equally spaced inputs.

Generally, the input points of a Gaussian process are assumed to be arbitrary. However, in certain application areas and problem settings, the input points are naturally equally spaced. Moreover, the assumption of equally spaced inputs can enable analytical tractability and provide valuable theoretical insights. The analytical solution derived in this study is particularly useful in such situations.

Specifically, the following points support the validity of the equally spaced input assumption in our study:
\subsection{Applicability to Equally Spaced Sampling in Practical Applications}
The results of this study are particularly useful when dealing with data sampled at equal intervals. For instance, in time series analysis, data are often acquired at regular intervals using sensors. Time series data such as temperature, stock prices, and audio signals are frequently sampled at equal intervals, and the analytical solution obtained in this study is applicable when applying Gaussian processes to these data. Furthermore, the assumption of equally spaced inputs can also be valid for modeling spatial fields using data obtained from sensors arranged at equal intervals on a spatial grid, such as meteorological observation data and soil moisture data. In addition, cases where the input data are inherently arranged in a regular grid, such as pixel data in digital images, are also possible. The assumption is not limited to these examples and can be reasonable or approximated as such in a wide range of situations.
\subsection{Enabling Theoretical Analysis}

By assuming equally spaced inputs, the kernel matrix can be diagonalized, and Fourier transform techniques can be applied to the relevant equations, which enabled us to derive an analytical solution for the mean squared error in this study. This analytical solution provides valuable theoretical insights into the behavior of Gaussian processes. In particular, since it is generally difficult to obtain analytical solutions for arbitrary input points, the analytical solution obtained under the specific condition of equally spaced inputs serves as a stepping stone for understanding the properties of Gaussian processes.
\subsection{Usefulness as a Benchmark for Numerical Methods and Approximations}

The analytical solution derived in this study serves as a benchmark for evaluating the accuracy of numerical methods and approximations for arbitrary input points. In general, approximation methods such as variational inference and Markov Chain Monte Carlo methods are used to compute Gaussian processes for arbitrary inputs. The accuracy of these approximations can be assessed by comparing them with the analytical solution in the case of equally spaced inputs.

\subsection{Limitations and Future Work}

It is important to acknowledge that the analytical solution in this study is derived under the condition of equally spaced inputs and is not directly applicable to cases with arbitrary input points. This is a limitation of the present study. However, as discussed above, there are application areas where the assumption of equally spaced inputs is reasonable, and the analytical solution also has theoretical significance.

Future work includes deriving analytical results for arbitrary input points without assuming equally spaced inputs. While this is generally a challenging problem, the analytical solution obtained in this study for equally spaced inputs can provide a foundation for further theoretical development. Another important direction for future research is to apply the results of this study to various application areas and improve the accuracy of analysis for equally spaced sampled data.

\begin{figure}[htb]
	\centering
	\includegraphics[width=12cm,clip]{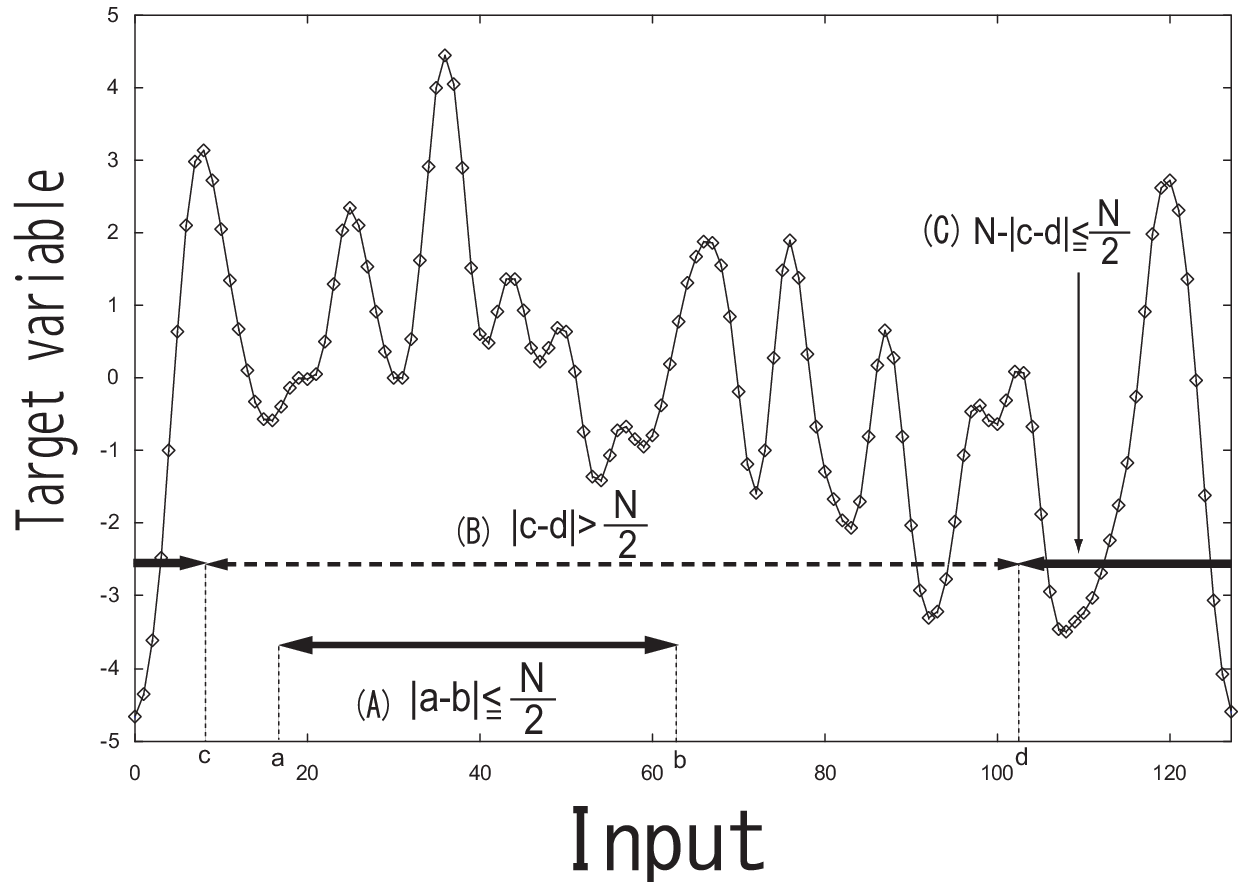}
\caption{Periodic boundary conditions are imposed on $N=128$ evenly spaced discrete input values. The target variable follows a Gaussian distribution. Under the periodic boundary condition, when the distance between two points exceeds $\frac{N}{2}$, the alternative distance between the two points spans across the $0$ point.}
\label{Gau-case}
\end{figure}

In this paper, we perform the following case distinctions to ensure that the model satisfies the periodic boundary conditions.
First, let $N$ denote the number of input elements. In making distinctions, as shown in FIG.\ref{Gau-case}(A), if the distance $|a - b|$ between two points does not exceed $\frac{N}{2}$, the distance is kept as is.
On the other hand, as shown in FIG. \ref{Gau-case}(B), if the distance $|c - d|$ between two points exceeds $\frac{N}{2}$, the distance between the two points is taken as $N - |c - d| $, as illustrated in FIG. \ref{Gau-case}(C).

In the study by Tsuzurugi and Eiho \cite{natural}, periodic boundary conditions were also used for natural image processing. The absence of noticeable artifacts in the output images further supports the validity of using periodic boundary conditions.
In this paper, we enable statistical mechanical calculations by using a model with periodic boundary conditions and one-dimensional input, as illustrated in FIG. \ref{Gau-case}. By applying the covariance matrix obtained under these modeled conditions to a Gaussian process, we derive the exact analytical solution of the Gaussian process. We then demonstrate that this analytical solution is consistent with the simulation results.
\section{Formulation}
In general, when the number of inputs is $N$, the input values of the Gaussian process $\mbx{x}_j(1 \leq j \leq N)$ are given in $D$ dimensions.
On the other hand, the target variable $t_j (1 \leq j \leq N)$, which is the output corresponding to the input values, is one-dimensional.
In this case, the dataset for training is defined as follows.
\begin{equation}
	\left\{
		\begin{array}{rl}
		\mbx{x}_j & \mbox{(\(1\leq j \leq N\))} \\
		t_j & \mbox{(\(1\leq j \leq N\))}
		\end{array}
	\right.\label{Gau-nyuryoku1}
\end{equation}
In the Gaussian process, it is assumed that the target variable $t_j$ already has noise superimposed, following a Gaussian distribution with a mean of $0$ and a variance of $b^2$. Additionally, when the unknown input data in $D$ dimensions is denoted as $\mbx{x}_{\bf{in}}$, its one-dimensional target variable is denoted as $t_{\bf{out}}$. In this case, the unknown dataset is defined as follows.
\begin{equation}
	\left\{
		\begin{array}{rl}
		\mbx{x}_{\bf{in}} &  \\
		t_{\bf{out}} & 
		\end{array}
	\right.
\end{equation}
In this paper, to further simplify the analysis, we model the inputs of the Gaussian process, as given in equation, as equally spaced discrete values $\mbx{j}={j}$ and assume periodic boundary conditions for the target values $\mbx{t}={t_j}$. Therefore, the possible values of the elements of the input vector $\mbx{j}$ are one-dimensional, and the training dataset is defined as follows.
\begin{equation}
	\left\{
		\begin{array}{rl}
		j & \mbox{(\(1\leq j \leq N\))} \\
		t_j & \mbox{(\(1\leq j \leq N\))}
		\end{array}
	\right.
\end{equation}
Let $N$ be an even number. Additionally, let the one-dimensional unknown input data be $j_{\bf{in}}$ and its target variable be $u_{\bf{out}}$. Then, the unknown dataset can be defined as follows.
\begin{equation}
	\left\{
		\begin{array}{rl}
		j_{\bf{in}} &  \\
		u_{\bf{out}} & 
		\end{array}
	\right.
\end{equation}
In a Gaussian process, both the true target variables without noise and the noise itself are given by Gaussian distributions. First, the probability distribution of the true target variables without noise, $\mbx{S} = {S_j}$, is given as follows.
\begin{equation}
	P(\mbx{S}) = \frac{1}{Z_{\mbx{S}}}\exp{\left(-\frac{1}{2}\mbx{S}^T A^{-1} \mbx{S}\right)}
	\label{Gau-seisei}
\end{equation}
\begin{equation}
	Z_{\mbx{S}} = (2\pi)^{\frac{N}{2}}|A|^{\frac{1}{2}}
\end{equation}
Here, $A$ is the covariance matrix, and to satisfy the periodic boundary conditions, its elements $A_{fj}$ are given by the following equations (\ref{Gau-nor}) and (\ref{Gau-abn}).
\begin{eqnarray}
& &A_{fj} = a^2 \exp [ -v^2|f-j|^2 ] \label{Gau-nor}\\
& &~~~~~~~~~~~~~~~~~~~~~~~~~~~~(|f-j| \le  \frac{N}{2})
\end{eqnarray}
\begin{eqnarray}
& &A_{fj} = a^2 \exp [ -v^2(N-|f-j|)^2 ]\label{Gau-abn} \\
& &~~~~~~~~~~~~~~~~~~~~~~~~~~~~(|f-j| >  \frac{N}{2})
\end{eqnarray}
Here, $a$ and $v$ are real numbers. The larger the value of $v$, the greater the correlation between neighboring target variables, leading them to take similar values. Conversely, when $v$ is small, the target variables tend to take independent values. The subscript $f$ is used for $A_{fj}$ instead of $i$ to avoid confusion with the imaginary unit $i$ used in the Fourier transform discussed later.
The probability distribution of the target variables with noise (degraded data) $\mbx{t} = {t_j}$ is given as follows,
\begin{equation}
	P(\mbx{t}|C) = \frac{1}{Z_{C}}\exp\left(-\frac{1}{2} \mbx{t}^T C^{-1} \mbx{t} \right)
	\label{Gau-noised}
\end{equation}
\begin{equation}
	Z_{C} = (2\pi)^{\frac{N^D}{2}}|C|^{\frac{1}{2}}
\end{equation}
$C$ is the covariance matrix, and its elements $C_{fj}$ are given by the following equation.
\begin{equation}
	C_{fj} = A_{fj} + B_{fj}
\end{equation}
Here, $B$ is the covariance matrix of the noise distribution, and its elements $B_{fj}$ are given by the following equation.
\begin{equation}
	B_{fj} = b^2 \delta_{f,j}
	\label{Gau-B}
\end{equation}
$b$ is a real number, and $\delta_{f,j}$ is the Kronecker delta, given by the following equation.
\begin{equation}
\delta_{f,j} = 	\left\{
		\begin{array}{rl}
		1, & f=j \\
		0, & f\neq j
		\end{array}
	\right.\label{Gau-delta}
\end{equation}
In this case, the noise is the difference between the degraded data $\mbx{t}$ and the true target variables $\mbx{S}$, and the probability distribution of the noise is given by the following equation.
\begin{eqnarray}
	P_{\bf{out}}(\mbx{t}|\mbx{S}) &=& \frac{1}{Z_{\bf{noise}}}\exp\left( -\frac{1}{2}(\mbx{t}-\mbx{S})^T B^{-1}(\mbx{t}-\mbx{S}) \right)\nonumber \\
	& &
	\label{Gau-noise}
\end{eqnarray}
\begin{equation}
	Z_{\bf{noise}} = (2\pi)^{\frac{N^D}{2}}|B|^{\frac{1}{2}}
\end{equation}
Here, the posterior probability function of the unknown data $u_{\bf{out}}$ can be calculated given the training target variables.
\begin{equation}
	P(u_{\bf{out}}|\mbx{t}, a^2, v^2, b^2) = \frac{P(u_{\bf{out}}, \mbx{t}|a^2, v^2, b^2)}{P(\mbx{t}|a^2, v^2, b^2)}
\end{equation}
The correlation between the unknown input ${j}_{\bf{in}}$ and the elements of the training input ${j}$ is given as follows.
\begin{equation}
	k({j}, {j}_{\bf{in}}) = a^2 \exp\left[ -v^2 ({j} - {j}_{\bf{in}})^2 \right]
\end{equation}
Furthermore, the following can be stated.
\begin{eqnarray}
	& &k_{\bf{in}}({j}_{\bf{in}}, {j}_{\bf{in}}) \nonumber \\
	&=& a^2 \exp\left[ -v^2 ({j}_{\bf{in}} - {j}_{\bf{in}})^2 \right]+b^2 \delta_{\bf{in}, \bf{in}}\nonumber \\
	&=& a^2 + b^2
\end{eqnarray}
Here, we define the vector ${\mbx{k}}$ with the element $k({j}, {j}_{\bf{in}})$ as follows.
\begin{eqnarray}
	\mbx{k}^T&=&\left\{k({1}, {j}_{\bf{in}}), \cdots, k({j}, {j}_{\bf{in}}), \cdots, k({N}, {j}_{\bf{in}})
				\right\} \nonumber \\
				&=&\{k_{f} \}
\end{eqnarray}
The mean of the unknown target variables is given by the following equation\cite{Williams-Gau}-\cite{ishii-Gau}.
\begin{eqnarray}
	 u_{\bf{out}}&=&\mbx{k}^T C^{-1} \mbx{t} \nonumber \\
	 &=&\sum_{{f}}\sum_{{j}}k_{{f}}C_{{f}{j}}^{-1} t_{{j}} \label{Gau-imamade} 
\end{eqnarray}
\section{Diagonalization of the Covariance Matrix}
By imposing discretization and periodic boundary conditions on the probabilistic model under consideration, we focus on cases where the covariance matrix becomes a symmetric circulant matrix, i.e., a translationally symmetric matrix. Under these conditions, the Fourier transform of $t_{j}$ becomes possible.
\begin{equation}
	\tilde{t}_{p} = \frac{1}{\sqrt{N}} \sum_{j} t_{j} e^{-i (pj)}
	\label{Gau-fourier}
\end{equation}
Here, $i$ is the imaginary unit. The inverse Fourier transform of Equation is given by the following equation.
\begin{equation}
	t_{j} = \frac{1}{\sqrt{N}} \sum_{p} \tilde{t}_{p} e^{i (pj)}
	\label{Gau-gyaku}
\end{equation}
The possible values of the elements of the vector $\mbx{p}$ are given as follows.
\begin{equation}
	0,\frac{2}{N}\pi, \frac{4}{N}\pi,\cdots,\frac{2(N-1)}{N}\pi
\end{equation}
In the case of a general two-dimensional array, the Fourier transform $\tilde{R}_{{h}{k}}$ of an arbitrary circulant matrix $R_{{f}{j}}$ is given by the following equation.
\begin{equation}
\tilde{R}_{{h}{k}}=\frac{1}{N}
\sum_{{f}}\sum_{{j}}e^{i({h}{j}-{k}{f})}
	R_{{f}{j}}\label{Gau-Fourie}
\end{equation}
Here, we consider the Fourier transform of the matrix elements commonly used as covariance matrices, given by Equations and. Due to the imposed periodic boundary conditions, we distinguish the cases where $0 \le l \le \frac{N}{2}$. When ${f}$ and ${j}$ denote the positions of the elements, the original Gaussian process \cite{Williams-Gau}-\cite{ishii-Gau} defines $A_{{f}{j}}$ as follows.
\begin{equation}
	A_{{f}{j}} = a^2 \exp[-\sum_{d=1}^D v^2_d |{f}^d-{j}^d|^2]
\end{equation}
Here, $d$ is not an exponent but an index of the input dimension. For simplicity, this paper considers the input dimension to be one-dimensional, hence using the equation with $D = 1$. The result of calculating $A_{{f}{j}}$ by distinguishing cases is given by the following equation.
\begin{eqnarray}
& &\tilde{A}_{{k}} \equiv \tilde{A}_{{k}{k}} = a^2 \sum_{l=0}^{\frac{N}{2}}e^{-v^2 l^2}\cos(l{k}) \nonumber \\
& &~~~~~+a^2 \sum_{l=\frac{N}{2}+1}^{N-1}e^{-v^2(N-l)^2}\cos(l{k})\label{wa-taikaku}
\end{eqnarray}
Similarly, by diagonalizing Equation using the Fourier transform, we obtain the following equation.
\begin{equation}
	\tilde{B}_{{p}} = b^2
	\label{Gau-B_p}
\end{equation}
Here, by applying the Fourier representation to Equation, we obtain the following equation.
\begin{equation}
	u_{\bf{out}} = \sum_{{q}} \tilde{k}_{-{q}}\tilde{C}^{-1}_{{q}} \tilde{t}_{{q}}
	\label{Gaurestored}
\end{equation}
In this paper, the mean of the unknown target variable $u_{\bf{out}}$ is used as the restored data. Since $u_{\bf{out}}$ is a real number, the imaginary part of Equation can be disregarded. $\tilde{k}{-{q}}$ is the complex conjugate of $\tilde{k}{{q}}$. Furthermore, since $\tilde{C}{{q}}$ is diagonalized by the Fourier transform, to obtain $\tilde{C}^{-1}{{q}}$, it is sufficient to take the reciprocal of the diagonal elements of $\tilde{C}_{{q}}$. These are the advantages of introducing the Fourier transform under periodic boundary conditions.
\section{Explanation of the Theory}
From here, we will briefly explain the method to calculate the mean squared error $E_1$ between the restored data $u_{\bf{out}}$ obtained in the previous section and the true value data $S_{{j}}$ without noise. In this chapter, variables with a hat symbol $\hat{~}$ are defined as unknown hyperparameters. We define $E_1$ as follows. Since $E_1$ is the mean squared error between the original data $S_j$ and the restored data $u_{\bf{out}}$, the smaller the value, the better the model's output. Although Equation represents the target variable derived from arbitrary inputs and typically takes continuous values, in practice, when calculating it using Equation below, we will use discrete values and replace the subscript of $u_{\bf{out}}$ with $j$.
\begin{equation}
	E_1=
		\left|\left|
		\frac{1}{N}\sum_{{j}}
			\left(
				S_{{j}} - u_{{j}}
			\right)^2
	\right|\right|
	\label{Gau-E_1}
\end{equation}
Here, $\left|\left| \cdot \right|\right|$ denotes the average with respect to the joint probability distribution $P(\mbx{t},\mbx{S})=P_{\bf{out}}({\mbx{t}}|{\mbx{S}})P({\mbx{S}})$.

The Fourier representation of Equation is given as follows.
\begin{equation}
		E_1=
		\frac{1}{N}\sum_{\mbx{k}}
		\left\|
			\Big(
				\tilde{S}_{\mbx{k}}
				-\tilde{u}_{\mbx{k}}
			\Big)
			\Big(
				\tilde{S}_{-{\mbx{k}}}
				-\tilde{u}_{-{\mbx{k}}}
			\Big)
		\right\|
		\label{Gau-FE_1}
\end{equation}
Since $P(\mbx{t},\mbx{S})$ is diagonalized in the Fourier representation, it can be easily computed.

Here, by setting as in the following equation,
\begin{equation}
	\tilde{A}^{-1}_{\mbx{k}}+\tilde{B}^{-1}_{\mbx{k}} = \tilde{C}'_{\mbx{k}}
	\label{Gau-oku}
\end{equation}
We obtain the following equation.
\begin{eqnarray}
		& &
		\left\|
			\Big(
				\tilde{S}_{\mbx{k}}
				-\tilde{u}_{\mbx{k}}
			\Big)
			\Big(
				\tilde{S}_{-{\mbx{k}}}
				-\tilde{u}_{-{\mbx{k}}}
			\Big)
		\right\|\nonumber \\
		&=&
		\frac{1}{Z}\int \int d\tilde{S}_{\mbx{k}} d\tilde{t}_{\mbx{k}}
		\left| \tilde{S}_{\mbx{k}} - \tilde{u}_{\mbx{k}}\right|^2
		\nonumber \\
		& &\times\exp
		\Biggr[-\frac{\tilde{C}'_{\mbx{k}}}{2}\left|
		\tilde{S}_{\mbx{k}} - \frac{\tilde{B}^{-1}_{\mbx{k}}}{\tilde{C}'_{\mbx{k}}}\tilde{t}_{\mbx{k}}
		\right|^2
		\nonumber \\
		& &~~~~~~~~~~~~
		 - \biggl\{
		\frac{\tilde{B}^{-1}_{\mbx{k}}}{2} - 
		\frac{\left(\tilde{B}^{-1}_{\mbx{k}}\right)^2}{2\tilde{C}'_{\mbx{k}}}\biggr\}
		|\tilde{t}_{\mbx{k}}|^2 \Biggr] \nonumber \\
		&=&\frac{1}{\tilde{C}'_{\mbx{k}}} 
		+\left(\frac{\tilde{B}^{-1}_{\mbx{k}}}{\tilde{C}'_{\mbx{k}}}
		-\hat{\tilde{A}}_{-\mbx{k}} \hat{\tilde{C}}^{-1}_{\mbx{k}}\right)^2
		\frac{1}{\tilde{B}^{-1}_{\mbx{k}} - \frac{\left(\tilde{B}^{-1}_{\mbx{k}}\right)^2}{\tilde{C}'_{\mbx{k}}}}
		\nonumber \\
		& &\label{Gau-kyuu43}
\end{eqnarray}

Therefore, when the following condition is satisfied,
\begin{equation}
	\hat{a} = a,~~~~\hat{b} = b,~~~~\hat{v} = v
	\label{Gau-jyouken}
\end{equation}
The mean squared error $E_1$ reaches the following minimum value.
\begin{equation}
E_{1_{min}} = \frac{1}{N} \sum_{{k}} \frac{1}{\tilde{C}'_{{k}}}
\end{equation}
Thus, the value of $E_{1_{\text{min}}}$ is,
\begin{eqnarray}
& &E_{1_{min}}=\nonumber \\
& &\sum_{{k}} \frac{1}{N \left(
    					\frac{1}{a^2 \sum_{{l}} e^{-v^2 {l}^2} e^{i({k} \cdot {l})}+a^2 \sum_{{l}} e^{-v^2 {(N-l)}^2} e^{i({k} \cdot {l})}}
    					+\frac{1}{b^2}
    					\right)}\nonumber \\
    					& &
    \label{Gau-E1min}
\end{eqnarray}
is given by.

Similarly, the mean squared error between the true target variable without noise and the noisy target variable before restoration can be obtained as follows.
\begin{equation}
	E_2=
		\left|\left|
		\frac{1}{N}\sum_{{j}}
			\left(
				S_{{j}} - t_{{j}}
			\right)^2
	\right|\right|
	=\frac{1}{N}\sum_{{k}}{b}^2 = b^2
	\label{Gau-E_2}
\end{equation}
\section{Results}
\begin{figure}[h]
	\centering
	\includegraphics[width=12cm,clip]{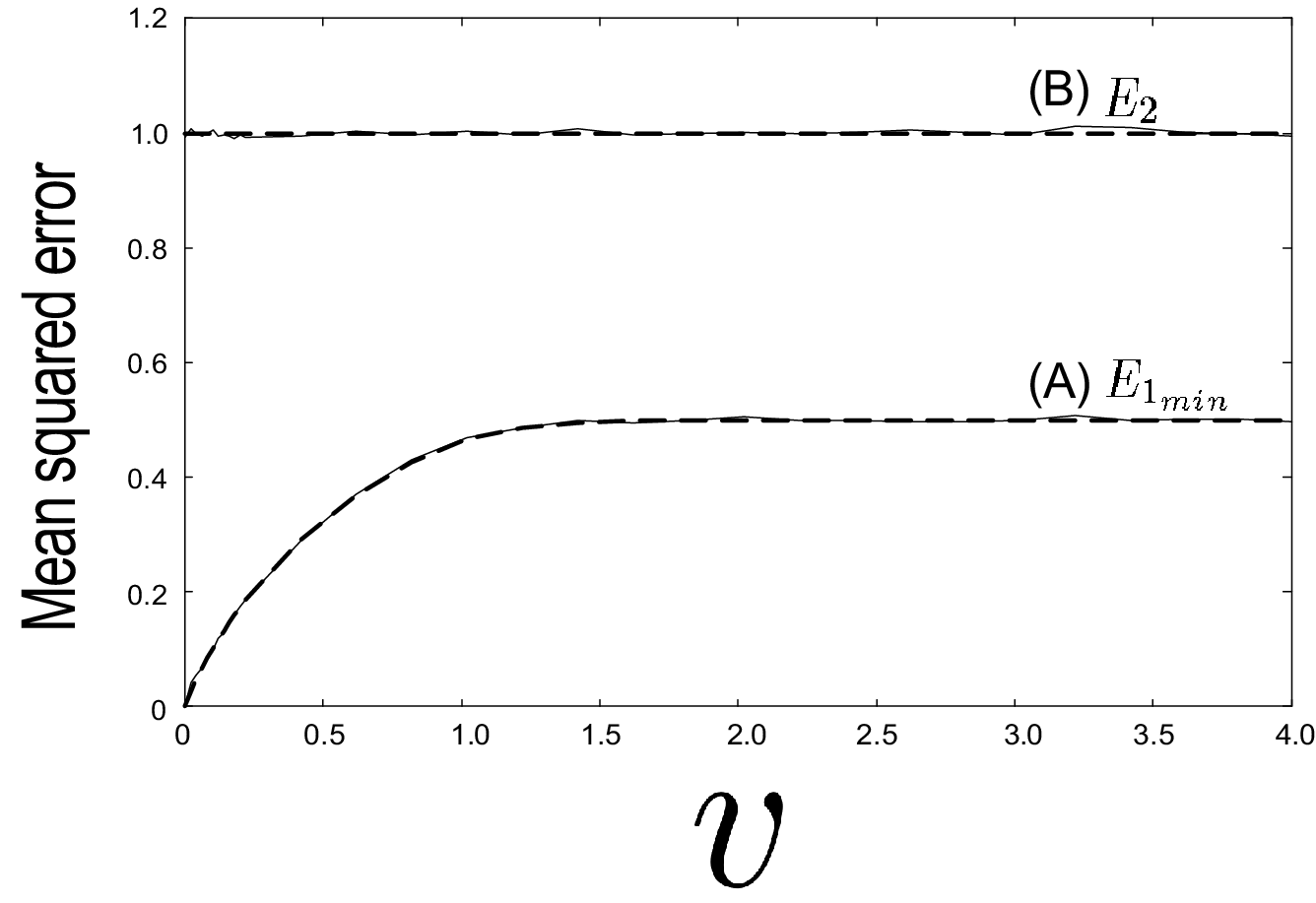}
\caption{
			Horizontal axis is $v$, Vertical axis is the mean squared error, in case of 
		$N=8192$, $a=1.0$, and $b=1.0$. 
		Dotted line (A) is the mean squared error of the oritical value, restored by 
			Eq.(\ref{Gau-E1min}). 
		Solid line (A) is the mean squared error by the numerical analisys 
			after restoration. 
		Dotted line (B) is the mean squared error before restoration 
			by Eq.(\ref{Gau-E_2}). 
		Solid line (B) is the mean squared error by the numerical analisys 
			before restoration. 
			}
\label{Gau-N=8192}
\end{figure}
From here, we present the analytical results of equations and.

FIG. \ref{Gau-N=8192} shows the mean squared error for $N=8192$ as an approximation of $N \rightarrow \infty$.  

Here, $a=1.0$ and $b=1.0$. The dashed line in (A) represents the theoretical value $E_{1_{\text{min}}}$, which is the mean squared error after restoration according to equation (\ref{Gau-E1min}). The solid line in (A) represents the mean squared error after restoration obtained from simulations. The dashed line in (B) represents the theoretical value $E_2$, which is the mean squared error before restoration according to equation. The solid line in (B) represents the mean squared error before restoration obtained from simulations. From FIG. \ref{Gau-N=8192}, it can be confirmed that the simulation results after restoration are in close agreement with the theoretical values. This suggests that for $N \rightarrow \infty$, the simulation results will perfectly match the theoretical values. Additionally, $N \rightarrow \infty$ is equivalent to the case where periodic boundary conditions are not applied.

\begin{figure}[htb]
	\centering
	\includegraphics[width=16cm,clip]{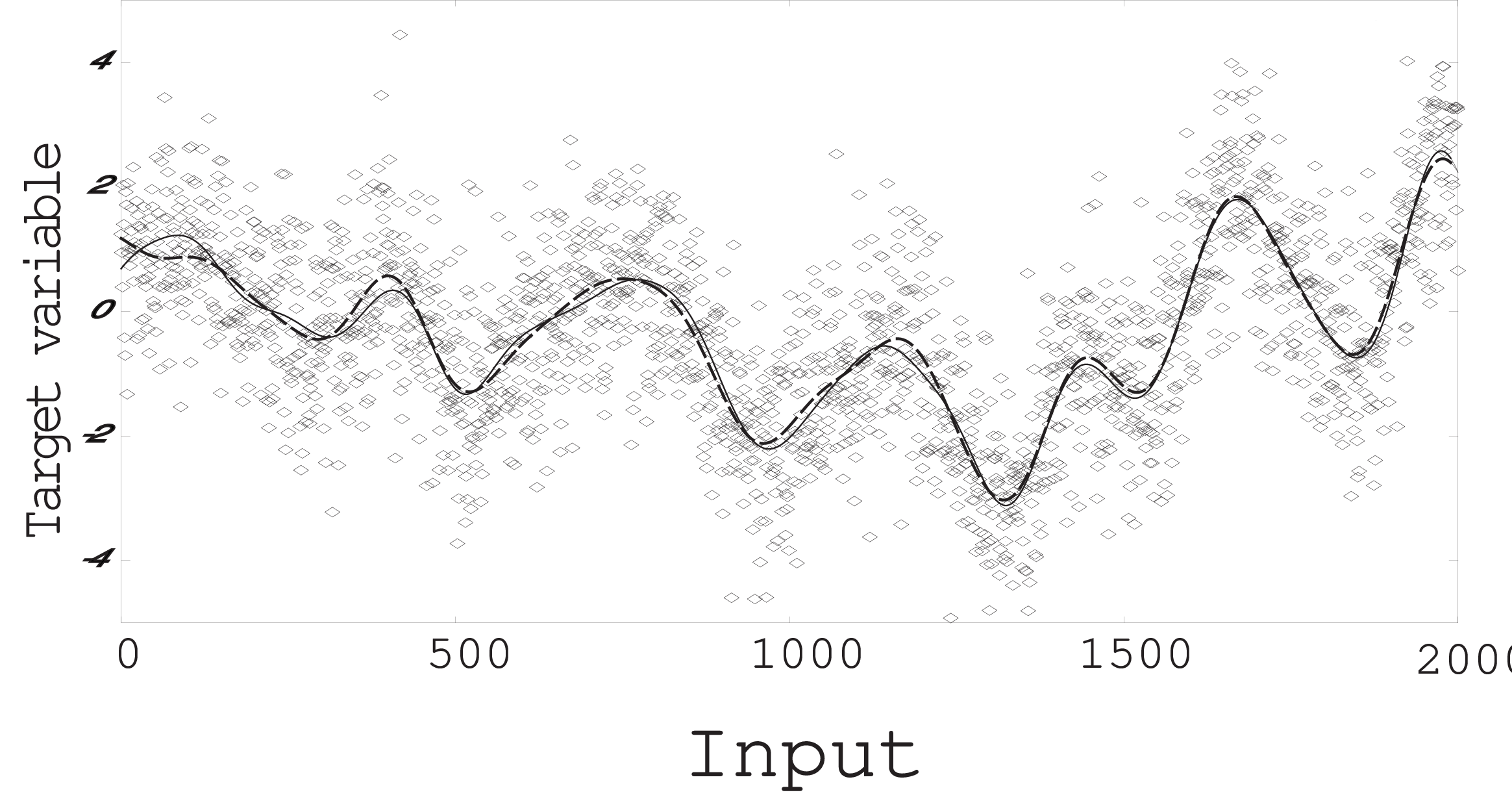}
\caption{Horizontal axis is Input, Vertical axis is Output, in case of 
		$v=0.08$, $N=8192$, $a=1.0$, and $b=1.0$. Solid line is Original data. 
		Dotted line is Restored data. Rhomboid is Distorted data. }
\label{Gau-N=8192v=001}
\end{figure}

Here, the behavior of the Gaussian process for $v=0.01$ is shown in FIG. \ref{Gau-N=8192v=001}.
For $v=0.01$, $N=8192$, $a=1.0$, and $b=1.0$. The horizontal axis represents the input, and the vertical axis represents the target variable. However, $N$ is $8291$, but to avoid making the figure difficult to read, the input is truncated at $2000$.
The solid line indicates the original data, the dashed line indicates the restored data, and the diamond markers indicate the degraded data.
From FIG. \ref{Gau-N=8192v=001}, it can be seen that the solid line representing the original data and the dashed line representing the restored data are in good agreement.
\section{Conclusion}
In this chapter, we performed precise diagonalization by case-separating a translationally symmetric matrix (translationally symmetric covariance matrix). Then, we applied the precisely diagonalized covariance matrix to a Gaussian process. As a result of applying it to the Gaussian process, it was found that the analytical solution and the simulation matched well in terms of the mean squared error before restoration ($E_2$) and after restoration ($E_1$).
\nocite{*}
\bibliography{sorsamp}

\begin{thebibliography}{99}
\bibitem{Bishop}
\newblock Christopher M. Bishop
\newblock ``Pattern Recognition and Machine Learning,''
\newblock Springer, Aug. 2006
\newblock ISBN-13    978-0387310732
\bibitem{Williams-Gau}
\newblock C.K.I. Williams and C.E.Ramussen
\newblock ``Gaussian process for regression,''
{Advanced in Neural Information Procesing System}, 8, 1996
\bibitem{Rasmussen}
\newblock C. E. Rasmussen
\newblock ``Evaluation of Gaussian process and other methods for non-linear regression,''
\newblock {Ph.D. thesis, Universitiy of Toronto}, 1996
\bibitem{Gibbs-Gau}
\newblock M. Gibbs and D.J.C MacKay
\newblock ``Efficient implementation of Gaussian process,''
{Technical report, Cavendish Laboratory, Cambridge, UK}, 1997
\bibitem{MacKay(1998)}
\newblock D.J.C.MacKay
\newblock ``Introduction to Gaussian process,''\\
\newblock In C.M.Bishop(Ed),Neural Networks and Machine Learning, pp.133--166. Springer, 1998
\bibitem{Williams}
\newblock C.K.I. Williams
\newblock ``Prediction with Gaussian process: from linear prediction and beyond,''
\newblock {In M.I. Jordan(Ed), Learning in graphical models, pp.599--621. MIT Press}, 1999
\bibitem{Smola-Gau}
\newblock A.J. Smola and P. Bartlett
\newblock ``Sparse greedy Gaussian process regression,''
{Advanced in Neural Information Procesing System}, 13, 2001
\bibitem{ishii-Gau}
\newblock T. Yoshioika, S. Oba, S. Ishii
\newblock ``Gaussian process regression using representing data,''
{Workshop Informaion-Based Induction Science}, pp.217--222, July. 2001
\bibitem{MacKay(2003)}
\newblock  D.J.C.MacKay
\newblock ``Information theory, inference and learning algorithms,''
\newblock {Cambridge University Press}, 2003
\bibitem{Cressie(1993)}
\newblock N.Cressie
\newblock ``Statistics for Spatial Data,''Wiley, 1993
\bibitem{Gau-juntuzu1}
\newblock J. Tsuzurugi, M. Okada
\newblock ``Statistical mechanics of Baysian image restorationunder spatially correlated noise,''
{Physical Review E}, vol.66, 066704, 2002
\bibitem{natural}
\newblock Tsuzurugi J, Eiho S, 
\newblock ``Image restoration of natural image under spatially correlated noise,''
\newblock {IEICE Transactions. Fundamentals}, Volume and Number: Vol.E92-A,No.3, Mar. 2009
\bibitem{Gau-Wahba}
\newblock G. Wahba
\newblock ``Spline models for observation data,''
\newblock {CBMS - NSF Regional Conference Series in Applied Mathematics}, 2006
\end{thebibliography}

\end{document}